\documentclass[aps,pra,reprint,superscriptaddress]{revtex4-2}
\usepackage{amsmath}
\usepackage{algorithm}
\usepackage{physics}
\usepackage{bbm}
\usepackage{amssymb}
\usepackage{slashed}
\usepackage{cancel}
\usepackage{graphicx}
\usepackage{mathrsfs}

\usepackage{graphicx}
\usepackage[normalem]{ulem}
\usepackage{color}
\usepackage{bm}
\usepackage{amsfonts}
\usepackage{booktabs}
\usepackage{multirow}
\def\mnn{\textcolor{black}}

\begin{document}


\title{Quantum machine learning with indefinite causal order}


\author{Nannan Ma}
\affiliation{Centre for Quantum Technologies, National University of Singapore, Singapore 117543, Singapore}

\author{P. Z. Zhao}
\affiliation{Centre for Quantum Technologies, National University of Singapore, Singapore 117543, Singapore}

\author{Jiangbin Gong}
\email{phygj@nus.edu.sg}
\affiliation{Centre for Quantum Technologies, National University of Singapore, Singapore 117543, Singapore}
\affiliation{Department of Physics, National University of Singapore, Singapore 117551, Singapore}
\affiliation{Joint School of National University of Singapore and Tianjin University, International Campus of Tianjin University, Binhai New City, Fuzhou 350207, China}


\date{\today}

\begin{abstract}
In a conventional circuit for quantum machine learning, the quantum gates used to encode the input parameters and the variational parameters are constructed with a fixed order. The resulting output function, which can be expressed in the form of a restricted Fourier series, has limited flexibility in the distributions of its Fourier coefficients.  This indicates that a fixed order of quantum gates can limit the performance of quantum machine learning.  Building on this key insight (also elaborated with examples),  
we introduce indefinite causal order to quantum machine learning.  Because the indefinite causal order of quantum gates allows for the superposition of different orders, the performance of quantum machine learning can be significantly enhanced. 
Considering that the current accessible quantum platforms only allow to simulate a learning structure with a fixed order of quantum gates, we reform the existing simulation protocol to implement indefinite causal order and further demonstrate 
the positive impact of indefinite causal order on specific learning tasks.   Our results offer useful insights into possible quantum effects in quantum machine learning.  
\end{abstract}

\maketitle

\section{introduction}

Arising from intrinsic quantum features such as superposition and entanglement,  quantum computation \cite{lloyd1996universal, divincenzo1995quantum, nielsen2002quantum} beating its classical counterparts is becoming a reality.  Numerous quantum algorithms \cite{shor1999polynomial, grover1996fast, montanaro2016quantum,liu2021rigorous} and experimental demonstrations \cite{peruzzo2014variational, kandala2017hardware, vandersypen2001experimental,arute2019quantum} have aimed to realize quantum advantages. Inspired by the capabilities of classical-computation-based machine learning \cite{carleo2019machine} across various fields \cite{merchant2023scaling,jumper2021highly,fawzi2022discovering}, the pursuit of a new era in machine learning founded on quantum computation \cite{biamonte2017quantum, cong2019quantum, schuld2019quantum} is natural and necessary to harness quantum advantages. In a typical quantum learning model, a variational parameterized quantum circuit \cite{mcclean2016theory,yuan2019theory, goto2021universal,mitarai2018quantum,havlivcek2019supervised, lloyd2020quantum, schuld2020circuit}, containing various quantum gates, serves a role analogous to the neural network in classical machine learning. The input vector $\boldsymbol{x}$ and variational parameters $\boldsymbol{\theta}$ are encoded as gate parameters. The output $y$ is represented by the expectation value of an observable measured on the final state. This learning structure gives rise to a mapping function $f_{\boldsymbol{\theta}}: \boldsymbol{x} \to y$, which is designed to approximate the ground truth relation from $\boldsymbol{x}$ to $y$. The architecture of a quantum circuit
fundamentally influences learning model's capability, 
akin to the role of structure in classical neural networks.

Given that different quantum gates are not commutable in general,
the ordering of quantum gates plays an important role in a quantum circuit and consequently in a quantum learning model.   Specifically, the learning capability of a quantum machine learning model 
can be analyzed in terms of the frequency spectrum and Fourier coefficients 
of a Fourier series that expresses the function $f_{\boldsymbol{\theta}}$.   The Fourier coefficients are determined by the multiplication of the quantum gates used in a learning model.  Now if, as in conventional quantum machine learning,  the quantum circuit has a fixed order of all involved quantum gates, then
the Fourier coefficients of the associated learning model lacks certain flexibility in response to changes in the variational parameters and therefore limiting the learning ability of the quantum model.  This key recognition motivates us to introduce indefinite causal order to relax this fixed-order limitation so as to achieve an enhanced learning ability.  The proposal of using indefinite causal order is also stimulated by the fact that it is a frontier topic that holds substantial advantages in quantum computation \cite{chiribella2013quantum,miguel2023enhancing,araujo2014computational} and quantum information \cite{salek2018quantum,goswami2020increasing,chiribella2021indefinite,guerin2016exponential}. 
Indeed, unlike classical physics with the order of operations being predefined, quantum physics permits novel causality \cite{thompson2018causal,oreshkov2012quantum,
oreshkov2019time}, thereby leading to a coherent control of
orders.
Up to now, indefinite causal order has been used to explore novel computational strategies and uncover more efficient algorithms for specific tasks \cite{zhao2020quantum,felce2020quantum,sazim2021classical,colnaghi2012quantum,feix2015quantum,ebler2018enhanced,procopio2019communication}. Moreover, the feasibility of implementing indefinite causal order has been demonstrated in experiments \cite{rubino2017experimental,rubino2021experimental,guo2020experimental}.
Our objective is to leverage the advantage of indefinite causal order so as to further unlock the potential of quantum machine learning algorithms. 

In particular, in this work we delve into the influence of causal orders on quantum learning protocols, revealing how the flexibility of orders with quantum properties relaxes the limitation arising from a fixed order in a quantum circuit in order to enhance the performance of quantum machine learning.
The improvement from the extended causal order can be investigated in terms of Fourier analysis.  The resulting advantage due to indefinite causal order is manifested as more flexible Fourier coefficients, as verified in some specific tasks. By embracing the adaptability of causal orders imbued with quantum characteristics, quantum machine learning, at least in the special examples discussed below, is seen to be more useful and powerful. 

This paper is organized as follows.  We first briefly explain the constraint imposed by a fixed causal order of quantum gates. 
Then, we construct a quantum machine learning model based on a more intricate causal order  and elaborate on 
how the improvement of the learning ability arises from indefinite causal order. Finally, we demonstrate how to simulate the indefinite causal order on a quantum circuit and further utilize two examples to verify the enhanced learning ability resulting from a more flexible causal order.

\section{impact of causal orders on quantum learning ability}

As we show, conventional quantum machine learning relies on parameterized quantum circuits. The structure of a quantum circuit for learning is hyper-parameterized, which typically remains fixed in designing a quantum machine learning protocol.  In a typical circuit, different gates generally do not commute with one another. This leads to a natural question: why one particular order should be implemented but the others.  The order of gates in a quantum circuit is hence expected to be a crucial factor in determining the learning ability of a quantum machine learning model. This implies that when expressing the learning model in the Fourier series, a chosen particular order of gates imposes a strong (but seems unnecessary) constraint on the resulting Fourier coefficients.   

To put the above-mentioned issue in a specific context, below we shall focus on the widely adopted encoding-variational structure, where the circuit consists of encoding blocks and  variational blocks. For simplicity, we consider a scenario involving a univariate function with the input variable $x \in \mathbb{R}$. 
In this learning structure, the encoding gate is in form of $g(x)=\exp(-ixH)$ with $H$ being a Hermitian generator, and a variational block $W(\boldsymbol{\theta})$ is comprised of $N$ gates $T^i(\boldsymbol{\theta}_i)$ with variational parameter $\boldsymbol{\theta}_i$, $i \in [1,...,N]$. The overall unitary evolution is then given by 
\begin{equation}  
U(x,\boldsymbol{\theta})=g(x)W(\boldsymbol{\theta})=g(x)T^N(\boldsymbol{\theta}_N)...T^1(\boldsymbol{\theta}_1).
\end{equation}
The output $f_{\boldsymbol{\theta}}(x)$ is set to be the expectation value of an observable $O$ such that 
\begin{equation}
		f_{\boldsymbol{\theta}}(x)=\bra{0}U^\dagger(x,\boldsymbol{\theta})OU(x,\boldsymbol{\theta})\ket{0},
\end{equation}
where $\ket{0}$ is the initial state.

To analyze the learning capability, we expand the function $f_{\boldsymbol{\theta}}(x)$ into a Fourier series \cite{schuld2021effect}. For this purpose, we diagonalize the generator $H\equiv V^{\dagger}\sum V$, where $\sum$ is a diagonal matrix with elements being the eigenvalues $\lambda_1,...\lambda_d$ of $H$. As such, the encoding gate yields $g(x)= V^{\dagger}\exp(-ix\sum)V$. Considering that $V$ is a unitary operator and thus can be absorbed into the neighboring variational block $W(\boldsymbol{\theta})$ and the observable $O$, the encoding gate can be simply taken as $g(x)=\exp(-ix\sum)$. The combination of this form of encoding gate $g(x)$ and the exact form of variational block $W(\boldsymbol{\theta})$ gives the $i$th element of the final state $\ket{\psi}=U(x, \boldsymbol{\theta})\ket{0}$,
\begin{equation}
\label{order1}
    \ket{\psi}_i=\sum_{j_1,...j_{N-1}=1}^{d}e^{-i\lambda_i}T^N_{ij_{N-1}}(\theta_N)\dots T^2_{j_2j_1}(\theta_2)T^1_{j_11}(\theta_1),
\end{equation}
\mnn{where $T^k_{mn}$ is an element of unitary matrix $T^k$.}
Thereafter, the output under a measurement yields
\begin{equation}
	f_{\boldsymbol{\theta}}(x)=\sum_{k,l \in [d]}e^{i(\lambda_k-\lambda_l)x}c_{kl}(\boldsymbol{\theta}),
\end{equation}
with 
\begin{align}
    c_{kl}(\boldsymbol{\theta})
=&\sum_{\boldsymbol{i},\boldsymbol{j}}(T^{1})^{\dagger}_{1i_{1}}(\theta_1)(T^{2})^{\dagger}_{i_{1}i_2}(\theta_2)\dots(T^{N})^{\dagger}_{i_{N-1}k}(\theta_N)O_{kl}
    \notag\\
    &T^N_{lj_{N-1}}(\theta_N)\dots T^2_{j_2j_1}(\theta_2)T^1_{j_11}(\theta_1).
\end{align}
Clearly then, the Fourier coefficients $c_{kl}(\boldsymbol{\theta})$ as functions of variational parameters $ \boldsymbol{\theta}$ are influenced by the adopted order of $N$ gates in a variational block. This already hints that a more intricate gate order corresponds to more complex Fourier coefficients, thereby potentially resulting in a better learning ability. 



In a conventional quantum circuit, the sequence of gates is predetermined and hence the causal order is fixed. Consequently, this circuit can be seen as a structure of slots with different gates to be inserted in. 
This kind of structure can be understood as a higher-order transformation \cite{wechs2021quantum} that maps certain quantum operators to other quantum operators, namely, a quantum supermap. Extending this, various permutations of gates can be combined in a distinguishable manner, thus defining the classical order. Put simply, for each slot the utilized gates and their specific order are known. The cumulative effect of such a classical order is a linear combination of distinct fixed orders. This allows us to achieve different orders simultaneously, but there is no coherence between these different orders. 
The situation changes if we introduce quantum superposition, where different orders coexist, implying the coherence between them.  Our hope is that this coherence can yield more sophisticated coefficient functions with respect to the variational parameters.  In the following, we take a specific example to illustrate three different causal orders and how to harness quantum coherence to improve the learning ability.

\begin{figure}[htb]
    \centering
\includegraphics[scale=0.25]{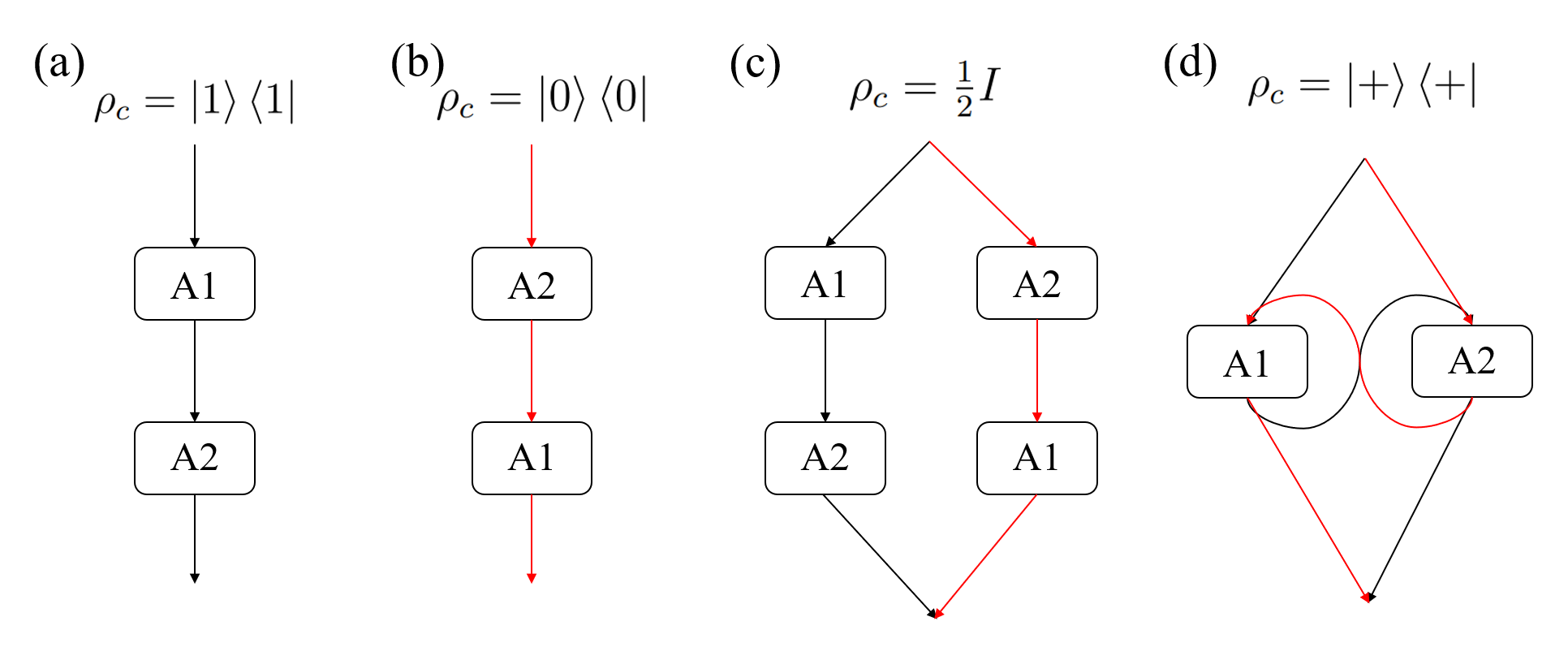}
    \caption{Illustration of different causal orders. (a) If $\rho_c$ is in $\ket{1}\bra{1}$, the causal order is $A_2\cdot A_1$. (b) If $\rho_c$ is in $\ket{0}\bra{0}$, the causal order is $A_1\cdot A_2$. (c) If $\rho _c$ is in the linear combination of two above state, i.e. in $I/2$ (classical 2 switch), the causal order is separable as the linear combination of two above definite orders. (d) If $\rho_c$ in the superposition (i.e. $\ket{+}\bra{+}$), the causal order is indefinite and cannot be separated into some definite causal orders, with the superposition of causal order (induced by $\ket{1}\bra{0}$ and $\ket{0}\bra{1}$).}
    \label{order}
\end{figure}

For a 2-switch structure depicted in Fig.~\ref{order}, the arrangements of two operators $A_1$ and $A_2$ define causal orders.
If the control system is in $\rho_c =\ket{1}\bra{1}$, the causal order is given by $A_2\cdot A_1$ (Fig.~\ref{order}(a)). Conversely, if $\rho_c =\ket{0}\bra{0}$, the causal order becomes $A_1\cdot A_2$ (Fig~\ref{order}(b)). 
Thereafter, for the control system $\rho_c=I/2$, the causal order yields a linear combination of the two definite causal orders, as shown in Fig~\ref{order}(c).
However, when $\rho_c$ is in a quantum superposition state $\ket{+}\bra{+}$ with $\ket{+}=(\ket{0}+\ket{1})/\sqrt{2}$, the causal order becomes indefinite and cannot be expressed as a combination of definite causal orders. Instead, the indefinite causal order now exhibits the coherence between two causal orders induced by items $\ket{1}\bra{0}$ and $\ket{0}\bra{1}$ (Fig~\ref{order}(d)).

Let us now generalize the 2-switch structure to the $N$-switch structure with the causal orders defined by the arrangements of $N$ operators $A_k$.
In this case, the control system basis state $\ket{\pi}$ signifies that the $N$ gates are deployed in the order $(\pi_1,...,\pi_N)$ and the resulting overall unitary gate is $\Tilde{A}_\pi = A_{\pi_N}...A_{\pi_1}$. The order here is encoded into the state $\ket{\pi}$ at the initialization stage of the control process, and there is no dynamical control involved. For different fixed orders $\pi \in\prod_{\mathcal{N}}$, we get different output function $f^{\pi}_{\boldsymbol{\theta}(x)}$ with the same frequency spectrum, but different coefficients
\begin{align}
    c^{\pi}_{kl}(\boldsymbol{\theta})=\sum_{\boldsymbol{i},\boldsymbol{j}}&(T^{\pi_1})^{\dagger}_{1i_{1}}(\theta_1)(T^{\pi_2})^{\dagger}_{i_{1}i_2}(\theta_2)\dots(T^{\pi_N})^{\dagger}_{i_{N-1}k}(\theta_N)O_{kl} \notag\\
    &T^{\pi_N}_{lj_{N-1}}(\theta_N)\dots T^{\pi_2}_{j_2j_1}(\theta_2)T^{\pi_1}_{j_11}(\theta_1),
\end{align}
where $\boldsymbol{\theta}$ is the variational parameter.
As a consequence, the new output with classical $N$-switch is in the form of the linear combination of different orders such that
\begin{equation}
f_{\boldsymbol{\theta}}(x)=\sum_{\pi}p_\pi\sum_{k,l\in [d]}c_{kl}^{\pi}e^{i(\lambda_k-\lambda_l)x},
\end{equation}
with $p_\pi \ge 0$ and $\sum_{\pi}p_\pi =1$. In this case, the spectrum is still the result of the difference of eigenvalues $\lambda_k-\lambda_l$ and remains the same as that in the fixed order case.
By contrast, the control system $\rho_c$ of a quantum $N$-switch structure includes not only diagonal elements $\ket{\pi}\bra{\pi}$, similar to the fixed orders case, but also non-diagonal elements $\ket{\pi}\bra{\pi^{\prime}}$, where  $\pi$  and $\pi'$ are two different orders. Therefore, the  coefficients appearing in the output function contain new terms induced by the non-diagonal elements of $\rho_c$, besides those in Eq.~(6) induced by the diagonal elements. \mnn{For instance, with $\pi$ representing the ascending order $W(\mathbf{\theta})^{\pi}=T^N(\theta_N)...T^1(\theta_1)$ and $\pi'$ as the descending order 
$W(\mathbf{\theta})^{\pi'}=T^1(\theta_1)...T^N(\theta_N)$,} the coefficient corresponding to $\ket{\pi^{\prime}}\bra{\pi}$ is given by
\begin{align}
c_{kl}^{\ket{\pi'}\bra{\pi}}(\boldsymbol{\theta})=&\sum_{\boldsymbol{i},\boldsymbol{j}}(T^{N})^{\dagger}_{1i_{1}}(\theta_N)(T^{N-1})^{\dagger}_{i_{1}i_2}(\theta_{N-1})\dots
   \notag\\
    &(T^{1})^{\dagger}_{i_{N-1}k}(\theta_1)
    O_{kl}T^N_{lj_{N-1}}(\theta_N)\dots
     \notag\\
    &T^2_{j_2j_1}(\theta_2)T^1_{j_11}(\theta_1).
\end{align}   
Therefore, the consequent output function is
\begin{equation}
    f_{\boldsymbol{\theta}}(x)=\sum_{\pi,\pi'}\rho_c^{\ket{\pi'}\bra{\pi}}\sum_{k,l\in [d]}c_{kl}^{\ket{\pi'}\bra{\pi}}e^{i(\lambda_k-\lambda_l)x},
\end{equation}
where $\rho_c^{\ket{\pi'}\bra{\pi}}$ is the coefficient of element $\ket{\pi'}\bra{\pi}$ in the control system state $\rho_c$. Here, the spectrum is still determined by $\lambda_k-\lambda_l$.

From the above discussion, it is clear that introducing controls of causal order keeps the frequency spectrum of the output mapping unchanged but alters the coefficients associated with respective frequencies. This flexibility of these coefficients allows an enhancement on the learning ability.
In contrast to classical control of causal order that  exploiting the linear combination of different orders, quantum control takes the flexibility a step further by introducing novel coefficients arising from the coherence of different orders, which do not exist in any fixed order. This property indicates a higher level of expressivity.

\section{Simulation of quantum causal order on a quantum circuit}

Although a typical $N$-switch has been demonstrated experimentally, the accessible quantum computation platforms, including relevant packages and quantum computers, are all with a fixed order. As such,  to explore the advantages of  quantum circuit with quantum causal order, we still rely on the conventional quantum circuit with a fixed order to 
proceed with the simulation, but will have to reform the existing method \cite{koska2021simulation}. 

The fundamental concept involves utilizing certain subsystems to assist in order control and subsequently discarding these subsystems. Let us now consider
the simulation of a quantum circuit with an $N$-switch with $N$ gates, denoted as $\{A_i\}$. The objective is to implement quantum causal order on the target system $\mathcal{H}^t$. Given that each order $\pi_i$, where $i \in \{1,...,N!\}$, is distinguishable, the bases $\ket{\pi}$ of an ancillary system $\mathcal{H}^{\alpha}$ represent these permutation orders. There are $N$ additional working target systems $\mathcal{H}^{t_1},...,\mathcal{H}^{t_{N}}$, each isomorphic to $\mathcal{H}^{t}$. Additionally, we require a control system $\mathcal{H}^c\otimes \mathcal{H}^{C_H}$ to manage the insertion of gates. In the control system, $\mathcal{H}^c$ is spanned by $\{\ket{k}^{c}\}^{N}_{k=1}$, with the basis $\ket{k}^c$ indicating the gates inserted into the current slot. Simultaneously, $\mathcal{H}^{C_H}$ is spanned by $\{\otimes_{i=1}^{N}\ket{\mathcal{K}_{i}}\}$, where $\mathcal{K}_{i}=1$ if the $i$th operator $A_i$ has been applied and $\mathcal{K}_{i}=0$ otherwise.

The primary control method is as follows: The overall initial state is given by
\begin{equation}
\ket{\psi_{0}}=\ket{\psi}^t\otimes\ket{\psi_\pi}^\alpha\otimes\ket{0}^c\otimes\ket{\emptyset}^{C_H}\bigotimes_{i=1}^{N}\ket{0}^{t_i},~ 	 \forall\psi,
\end{equation}
\mnn{where $\ket{\psi}^t$ is the initial target state and $\ket{\psi_\pi}^\alpha=\sum_{i}c_i\ket{\pi_i}^\alpha$ represents the superposition of different order.}
We employ control operators $U_1$ between the initial state and the first slot, $U_{n+1}$ between the $n$th and $(n+1)$th slots for $n \in \{1,...,N-1\}$, and $U_{N+1}$ between the $N$th slot and the final state. According to the order state $\ket{\pi}$ of the ancillary system, control operator $U_1$ transforms the initial state $\ket{0}$ of $\mathcal{H}^c$ to $\ket{\pi(1)}$. Control operator $U_{n+1}$ changes the state $\ket{\pi(n)}$ of $\mathcal{H}^c$ to $\ket{\pi(n+1)}$ and flips the state $\ket{\mathcal{K}_{\pi(n)}}$ of $\mathcal{H}^{C_H}$ from $\ket{0}$ to $\ket{1}$. Control operator $U_{N+1}$ changes the state $\ket{\pi(N)}$ of $\mathcal{H}^c$ to $\ket{0}$ and flips the state $\ket{\mathcal{K}_{\pi(N)}}$ of $\mathcal{H}^{C_H}$ from $\ket{0}$ to $\ket{1}$. Here, $\pi(n)$ represents the gate inserted in slot $n$ with the order $\pi$. Following the control operator, a Controlled-SWAP operator swaps the target system $\mathcal{H}^{t}$ and the additional working system $\mathcal{H}^{t_{\pi(n)}}$ as indicated by the state $\ket{\pi(n)}$ of the system $\mathcal{H}^c$. At each slot, gate $A_k$ is applied to the $k$th additional working system $\mathcal{H}^{t_k}$ to ensure that the target state undergoes the gate $A_{\pi(n)}$. Subsequently, another Controlled-SWAP is applied to swap the target state back to the target system $\mathcal{H}^t$.

In the above discussion, all subsystems are assumed to have proper dimensions, but this condition is not realistic in practice. For example, the $N!$-dimensional ancillary system represents the different orders of the $N$-switch structure. However, in a quantum circuit, this system should consist of $n$ qubits, as $2^n \ge N!$, and we need to take into account the redundant $2^n - N!$ dimensions.
To implement this simulation method, we need to provide the realistic and feasible structure of the simulation circuit. That is, in an $N$-switch structure, there are $N!$ different orders, and $N$ gates are inserted in these orders. These characteristics necessitate the consideration of some redundancies in these systems. The ancillary system $\mathcal{H}^{\alpha}$ comprises $N_\alpha=\lceil \log(N!) \rceil$ qubits. The first $N!$ basis vectors, forming the effective space $E$, represent the permutation orders, while the rest constitute a redundancy space $R$. In the control system, $\mathcal{H}^{c}$ contains $N_c=\lceil \log(N) \rceil$ qubits. The first $N$ basis vectors, denoted as the effective space $c_E$, indicate one of $N$ gates to be inserted, and the rest form the redundancy space $c_R$. Additionally, $\mathcal{H}^{C_H}$ is composed of $N$ qubits. The target system consists of $N_t$ qubits, and the $N$ additional working systems are all isomorphic to the target system, requiring $N N_t$ qubits. Therefore, a total of $N_t(N+1)+N_\alpha+N+N_c$ qubits are necessary in the quantum circuit. These redundancy spaces $R$ and $c_R$ (ignored in previous work) should be treated appropriately. Taking into account the redundancy spaces, we clarify the control unitary operators $U_i$,

\begin{widetext}
   \begin{gather}
   U_1=\mathbbm{1}^t\otimes\mathbbm{1}^{C_H}\otimes\bigg(\sum_{\pi\in E}\ket{\pi}\bra{\pi}^\alpha\otimes\sum_{i\in cE}\ket{\pi(1)+i}\bra{i}^c+\sum_{\pi\in E}\ket{\pi}\bra{\pi}^\alpha\otimes\sum_{i\in cR}\ket{i}\bra{i}^c+\sum_{\pi\in R}\ket{\pi}\bra{\pi}^\alpha\otimes\sum_{i}\ket{i}\bra{i}^c\bigg)\notag,\\
    U_n=\left(\mathbbm{1}^{tC_H}\otimes \text{SHIFT}^{(c,\alpha)}\right)\left(\mathbbm{1}\otimes \text{ExUnion}^{(C_H,c)}\right), \quad n\in\{2,...,N\},\notag \\
	U_{N+1}=\left(\mathbbm{1}^{tC_H}\otimes \text{FINAL}^{(\alpha,c)}\right)\left(\mathbbm{1}\otimes
    \text{ExUnion}^{(C_H,c)}\right), 
    \end{gather}
with some functional components 

    \begin{gather}
	\text{ExUnion}^{(C_H,c)}=\sum_{i\in cE}\ket{i}\bra{i}^c\otimes\sum_{K}\ket{K\bcancel{\cup}\{i\}}\bra{K}^{C_H}+\sum_{i\in cR}\ket{i}\bra{i}^c\otimes\mathbbm{1}^{C_H},\notag \\
	\text{SHIFT}^{(c,\alpha)}=\sum_{j\in cE}\sum_{\pi \in E}\ket{\pi}\bra{\pi}^\alpha\otimes\ket{\pi(\pi^{-1}(j)+1)}\bra{j}^c+\left(\sum_{j\in cE}\sum_{\pi \in R}+\sum_{j\in cR}\sum_{\pi \in E}+\sum_{j\in cR}\sum_{\pi \in R}\right)\ket{\pi}\bra{\pi}^\alpha\otimes\ket{j}\bra{j}^c,\notag\\
    \text{FINAL}^{(\alpha,c)}=\left(\sum_{\pi\in E \& j\in cE}\ket{\pi}\bra{\pi}^\alpha\otimes\ket{j}\bra{\pi(N)+j}^c+\sum_{\text{otherwise}}\ket{\pi}\bra{\pi}^\alpha\otimes\ket{j}\bra{j}^c\right),
    \end{gather}
where $ K\bcancel{\cup}\{i\}=K\cup{i}$ if $ {i}\not\subset K$, else $K\bcancel{\cup}{i}=K\setminus\{i\}$. At each time slot, the Controlled-SWAP gate with $A_k$ acting on the additional working systems constructs the gate the $ \tilde{A}' $, resulting in
\begin{equation}
	\tilde{A}' =\left(\sum_{k\in cE}\ket{k}\bra{k}^c\otimes \text{SWAP}(t,t_k)+\sum_{k\in cR}\ket{k}\bra{k}^c\otimes I\right)\left(\bigotimes_{k=0}^{N-1}A_k\right)\left(\sum_{k\in cE}\ket{k}\bra{k}^c\otimes \text{SWAP}(t,t_k)+\sum_{k\in cR}\ket{k}\bra{k}^c\otimes I\right)
\end{equation}
with the SWAP gate that swaps the states of two $d$-dimensional quantum systems given by
\begin{equation}
	\text{SWAP}(a,b)=\sum_{i,j=0}^{d-1}\ket{i}\bra{j}^a\otimes\ket{j}\bra{i}^b.
\end{equation}
With such a construction, the overall quantum circuit for the $N$-switch structure is apparent, and the final state $\ket{\psi_f}$ is given by
\begin{equation}
\ket{\psi_f}=U_{N+1}\tilde{A}'U_{N}\ldots U_{2}\tilde{A}'U_{1}\ket{\psi_0}.
\end{equation}
In this final state $\ket{\psi_f}$, the superposition of $N!$ permutation orders is achieved with the superposition of the ancillary system state $\ket{\psi_\pi}^\alpha$. It is noted that there are some redundant elements from the bases $\ket{\pi}^\alpha$, where $\pi\in R$. This redundant part can be discarded by setting the ancillary part in the observable as a Hermitian operator existing in the effective space $E$.

\end{widetext}

\section{Verification of improvement from indefinite causal order}

After having illustrated the simulation method in the above section, let us now take two examples to showcase the benefits of a more flexible causal order.
One is a quantum learning model with a 2-switch structure, where the output function with indefinite causal order performs better than that with fixed and classical orders. 
The other one is applying a 3-switch structure to a binary classification task, where the classical order allows an improved accuracy
and the quantum order permits a higher-level improvement, compared with the fixed-order structure. 

\subsection{Quantum learning model with 2-switch}

Consider a 
simple quantum circuit with a 2-switch structure. In this learning structure, we take the initial state as $\ket{0}$, the input as $x$ (encoded by the rotation angle of a $R_X$ gate) and the variational parameter as $\theta$ (encoded by the rotation angle of a $R_Z$ gate). The observable is $\sigma_z$ and the output is the expectation value of this observable, denoted by $y$.
For this setup, there exist two causal orders, $R_Z(\theta)R_X(x)$ and $R_X(x)R_Z(\theta)$. With \mnn{the first order $R_Z(\theta)R_X(x)$}, the output state is $R_Z(\theta)R_X(x)\ket{0}$ and then the final mapping function is $\cos x$. We can find that the variational parameter $\theta$ disappears and hence this structure has no learning ability based on the fact we cannot adjust the variational parameter $\theta$ to fit the target function (not $\cos x$). 
For \mnn{the second order $R_X(x)R_Z(\theta)$}, the output function is still $\cos x$ and thus this circuit has no learning ability either. 

We now add an order control system $\mathcal{H}^c$ with $\ket{0}^c$ and $\ket{1}^c$ indicating the orders $R_Z(\theta)R_X(x)$ and $R_X(x)R_Z(\theta)$ respectively.
In this case, the initial state of whole system is $\ket{+}^c\otimes\ket{0}$. In the classical 2-switch structure, the whole observable reads  $I^c\otimes\sigma_z$ and the output is the linear combination of two different orders after tracing out the control system, thereby the output being still $\cos x$. As a result, the classical control version of this structure is unable to learn a general target function. However, 
in the quantum case, the coherence of two different orders is introduced into the learning model. This leads to the whole observable being  $\ket{+}\bra{+}^c\otimes\sigma_z$. It is straightforward that the coherence of different orders is from the items $\ket{0}\bra{1}^c$ and $\ket{1}\bra{0}^c$ in the control system. From this structure, a new function appears, that is  $[(3+\cos \theta)\cos x+1-\cos \theta]/4$. Clearly, the variational parameter $\theta$ reemerges and \mnn{an additional item with zero frequency} appears in this Fourier series. 
	
This structure can be made more complex by replacing $R_Z(\theta)$ with a general unitary gate involving three variational parameters such that $R_Z(\theta) \to U(\theta,\phi,\lambda)$.
 For the order $U(\theta,\phi,\lambda)R_X(x)$, the output function is given by $\cos \theta \cos x-\sin \theta\sin \lambda\sin x$. It is obvious that the variational parameter $\phi$ disappears and there is no constant item. 
 For the $R_X(x)U(\theta,\phi,\lambda)$, the output function reads $\cos\theta\cos x-\sin\theta\sin\phi\sin x$, which is similar to the above one except for the replacement between $\phi$ and $\lambda$ so that there is no constant item with any definite order and their mixture (namely,  classical control of order). But with the coherence of these two different orders, a new item emerges. It is $(1+\sin\phi\sin\lambda)\cos\theta-\cos\phi\cos\lambda+(\sin\phi-\sin\lambda)\sin\theta\sin x+[(1-\sin\phi\sin\lambda)\cos\theta+\cos\phi\cos\lambda]\cos x$ arising from the interference between two orders $\ket{0}\bra{1}$ and $\ket{1}\bra{0}$. The constant item appears, and all variational parameters can be used to optimize this learning model.

\subsection{Improvement in a classification task}

We have demonstrated the improvement of learning ability by analyzing  the output function.
In the following, we perform a binary classification task using a single qubit circuit with a two-dimensional dataset to showcase this learning advantage in practical scenarios.
Here, we use the package QISKIT \cite{Qiskit} to construct the quantum circuit and then train it with COBYLA so as to get optimal parameters.

The input is a 2D vector $x=[x_1, x_2]$ with $x_1, x_2 \in [-1,1]$. 
The distribution of the dataset is as follows: a square with a width of 2 is divided into two parts by a circle with a radius of $\sqrt{2/\pi}$. The samples inside (outside) the circle with label $y=-1$ ($1$) are marked as green (blue) points. The decision function is 
\begin{equation}
    y=\text{sgn} \left [(x_1)^2+(x_2)^2-\frac{2}{\pi}\right ].
\end{equation}
In this learning task, we utilize a single qubit system to perform the computation. The model includes a $R_Z$ gate for encoding the variable $x_1$, a $R_Y$ gate for encoding the variable $x_2$, and a unitary gate $U(\phi,\theta,\omega)$ containing three trainable parameters $[\phi,\theta,\omega]$. The observable is taken as $\sigma_z$. As the output, denoted as $e \in [-1,1]$, is the expectation value, we define the classification criterion as
\begin{equation}
	y=
	\begin{cases}
		-1,  ~~~e\leq0,\\
		1, \quad ~~e>0.
	\end{cases}
\end{equation}
To provide a clearer comparison, we simplify the setup by using only one layer that contains three gates mentioned above, without repeating it. The training dataset consists of 200 sample points. 
	
\begin{figure}[b]
    \centering
    \includegraphics[scale=0.4]{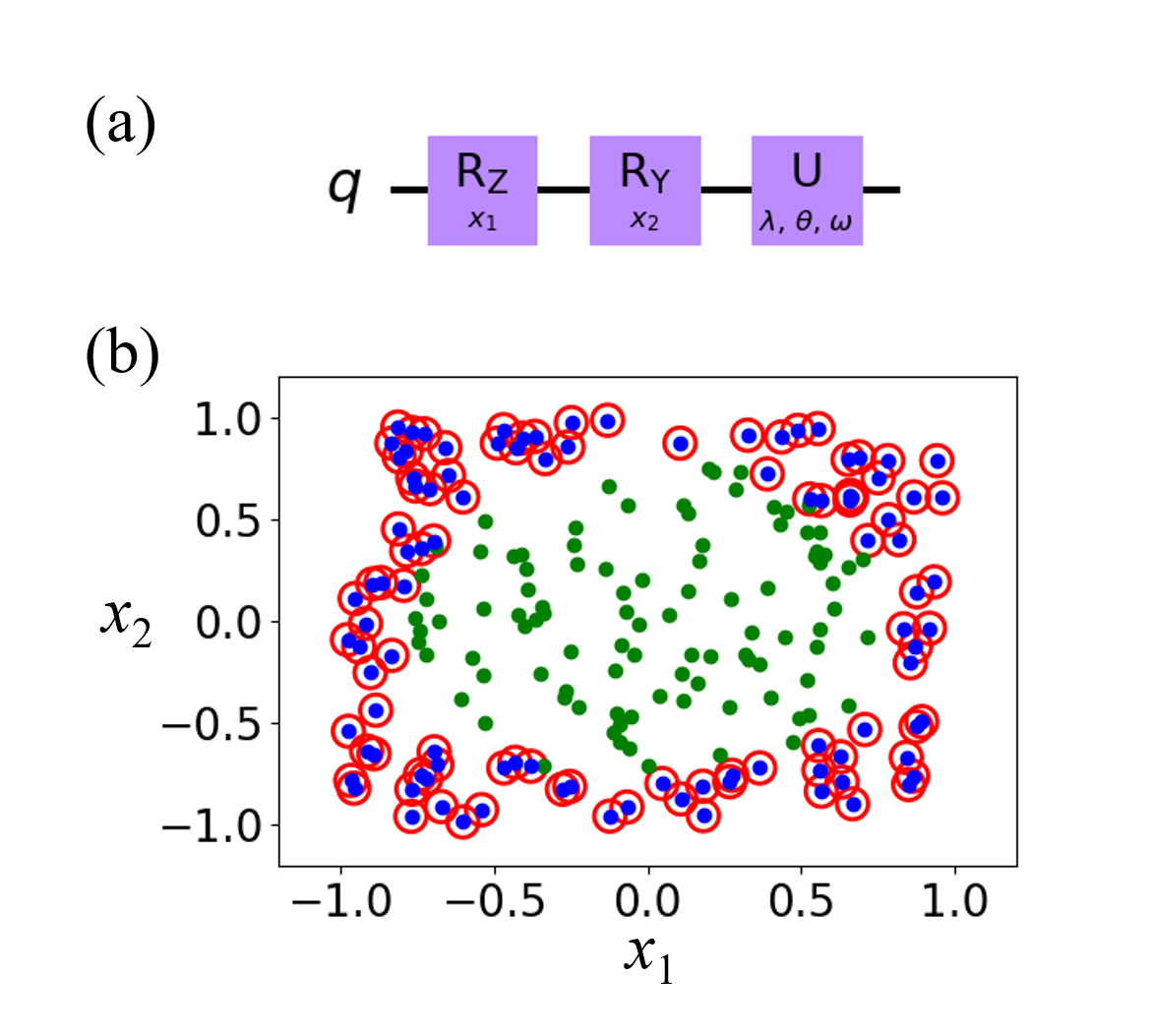}
    \caption{Circuit with the fixed order and the classification result for two dimensional binary classification task. (a) The circuit with fixed order. The first two gates $R_Z$ and $R_Y$ are used to encode the two dimensional input vector $[x_1, x_2]$, and the unitary gate with three training parameters $\{\lambda,\theta, \omega\}$ is as the variational part. (b) The
  classification result of 200 randomly selected samples. The wrongly classified samples are marked with red circles. The accuracy is $50\%$.}
    \label{fix}
\end{figure}

We start with the fixed order $U_{\text{overall}}=U(\phi,\theta,\omega)R_Y(x_2)R_Z(x_1)$, as shown in Fig.~\ref{fix}(a). The accuracy of classification for such structure is $50\% $ with the optimal parameters vector $[\lambda,\theta, \omega]=[1.6623, 0.8838, 1.5971]$. Then, we test this obtained model. The testing result is shown in Fig.~\ref{fix}(b), where all input points are  recognized as green. The result indicates that the structure lacks the ability to learn the target pattern. 

\begin{figure*}[htbp]
    \centering
    \includegraphics[scale=0.55]{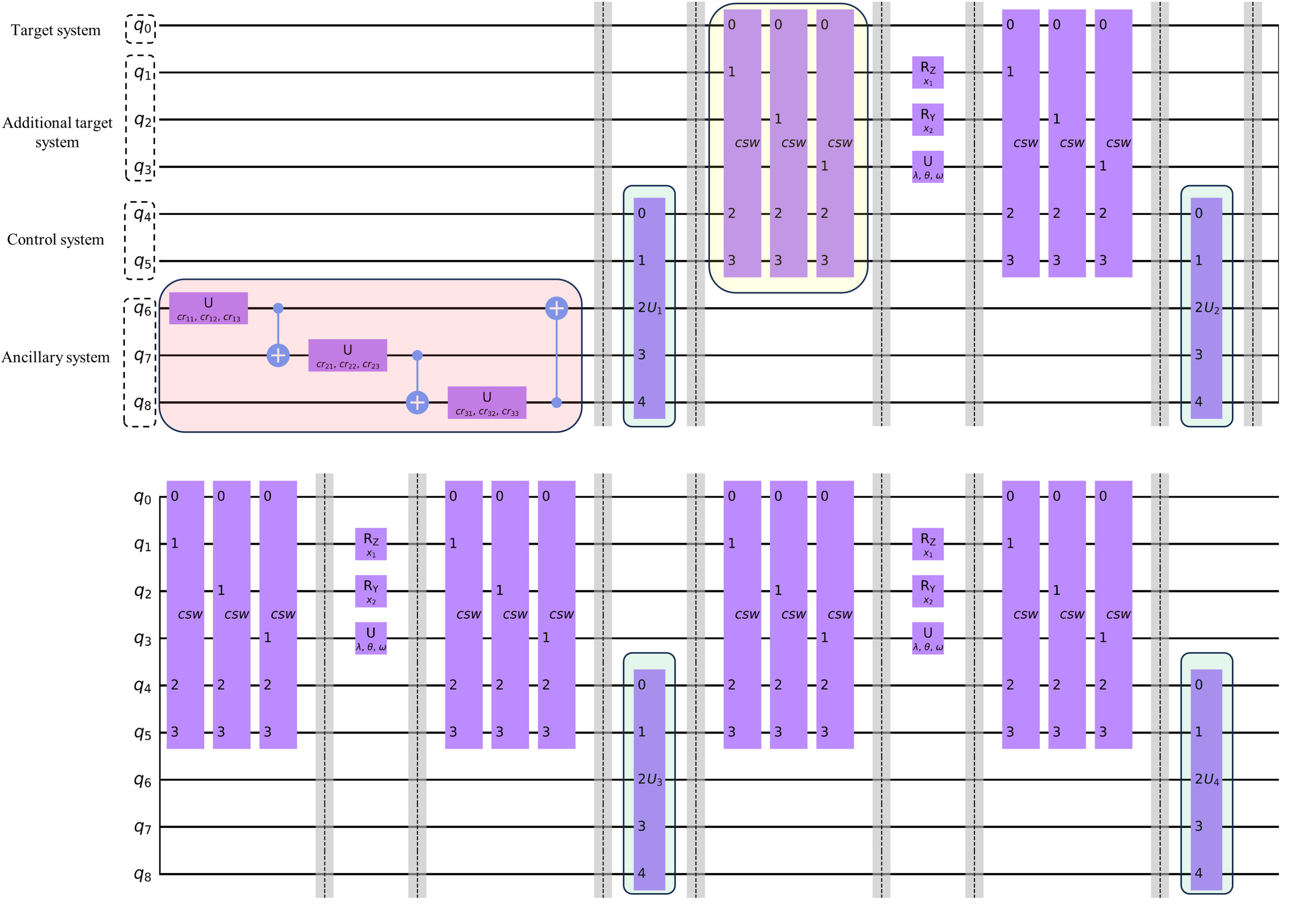}
    \caption{Circuit diagram for a 3-switch system utilizing 9 qubits. It depicts a 3-switch circuit for introducing classical control of causal order to actual gates $U(\phi,\theta,\omega)$, $R_Y(x_2)$, and $R_Z(x_1)$. The circuit comprises 9 qubits, with 3 qubits ($q_6, q_7, q_8$) allocated for the ancillary system $\mathcal{H}^{\alpha}$, 2 qubits ($q_4, q_5$) for the control system $\mathcal{H}^{c}$, and 4 qubits for the target system $\mathcal{H}^{t}$ and additional working target system $\mathcal{H}^{t_1t_2t_3}$. The flexibility in the ancillary system allows leveraging the superposition of different orders. The red block represents three unitary gates on the ancillary qubits with 9 trainable parameters $[cr_{ij}]$. Additionally, CNOT gates induce entanglement between different qubits in the ancillary system. Internal operations $U_1$, $U_2$, $U_3$ ,$U_4$ (blue blocks) connect the ancillary and control systems and transfer the control system's state based on the order state. The control system's basic states determine which actual gate is applied. The yellow block shows three control swap (csw) gates that swap target and additional target qubit systems based on the control system's state. Three control swaps are needed to construct the overall swap operation.}
    \label{nswitch}
\end{figure*}

Then, we introduce the circuit setup for simulating classical and quantum 3-switch with $U(\phi,\theta,\omega)$, $R_y(x_2)$ and $ R_z(x_1)$, the structure of which is shown in Fig.~\ref{nswitch}. 
The circuit consists of 9 qubits, where $ n_\alpha=3 $ qubits $(q_6, q_7, q_8)$ are allocated for the ancillary system $ \mathcal{H}^{\alpha}$, $ n_c =2 $ qubits $(q_4, q_5)$ are allocated for the control system $\mathcal{H}^{c}$, and qubit $q_0$ is allocated for the target $\mathcal{H}^{t}$. There is an additional working target system $(q_1,q_2,q_3) \in \mathcal{H}^{t_1t_2t_3}$. 
Considering that it is unnecessary to record which gate is used in our simulation, there is no system $\mathcal{H}^{C_H}$. 
For the pink block in Fig.~\ref{nswitch}, we place three unitary gates on the $ n_\alpha $ qubits, resulting in 9 trainable parameters $[cr_{ij}], i,j=1,2,3$. 
Besides, we add three CNOT gates into this block so as to produce entanglement between different qubits. This makes the states in $ \mathcal{H}^\alpha $ flexible enough to represent different distributions of 6 possible orders, labelled by the first six basic states, from $\ket{000}^{\mathcal{H}^\alpha}$ to $\ket{101}^{\mathcal{H}^\alpha}$, in this space.
Denoting $R_Z(x_1)$ as gate 0, $R_Y(x_2)$ as gate 1 and $U(\phi,\theta,\omega)$ as gate 2, then the sequences of the 6 possible orders are given by $\{[0,1,2],[0,2,1],[1,0,2],[1,2,0],[2,0,1],[2,1,0]\}$. 
The internal operations $U_1$, $U_2$, $U_3$ and $U_4$ in the circuit diagram are depicted as blue blocks. These operations is used to connect the ancillary and control systems, and to change the state of the control system based on the states in the ancillary system. The first three basis states $\ket{00}^{\mathcal{H}^c}$, $\ket{01}^{\mathcal{H}^c}$ and $\ket{10}^{\mathcal{H}^c}$ of the control system indicate the actual gate applied to the target state in the subsequent steps, and the last basis state $\ket{11}^{\mathcal{H}^c}$ is a redundancy. This mechanism allows for the control of the gate operations based on the specific order being processed. \mnn{The yellow block comprises three elementary controlled-swap gates, employed to swap the target system with one of three additional target qubit systems, effectively constructing an aforementioned Controlled-SWAP gate.}

\begin{figure}[htb]
    \centering
    \includegraphics[scale=0.25]{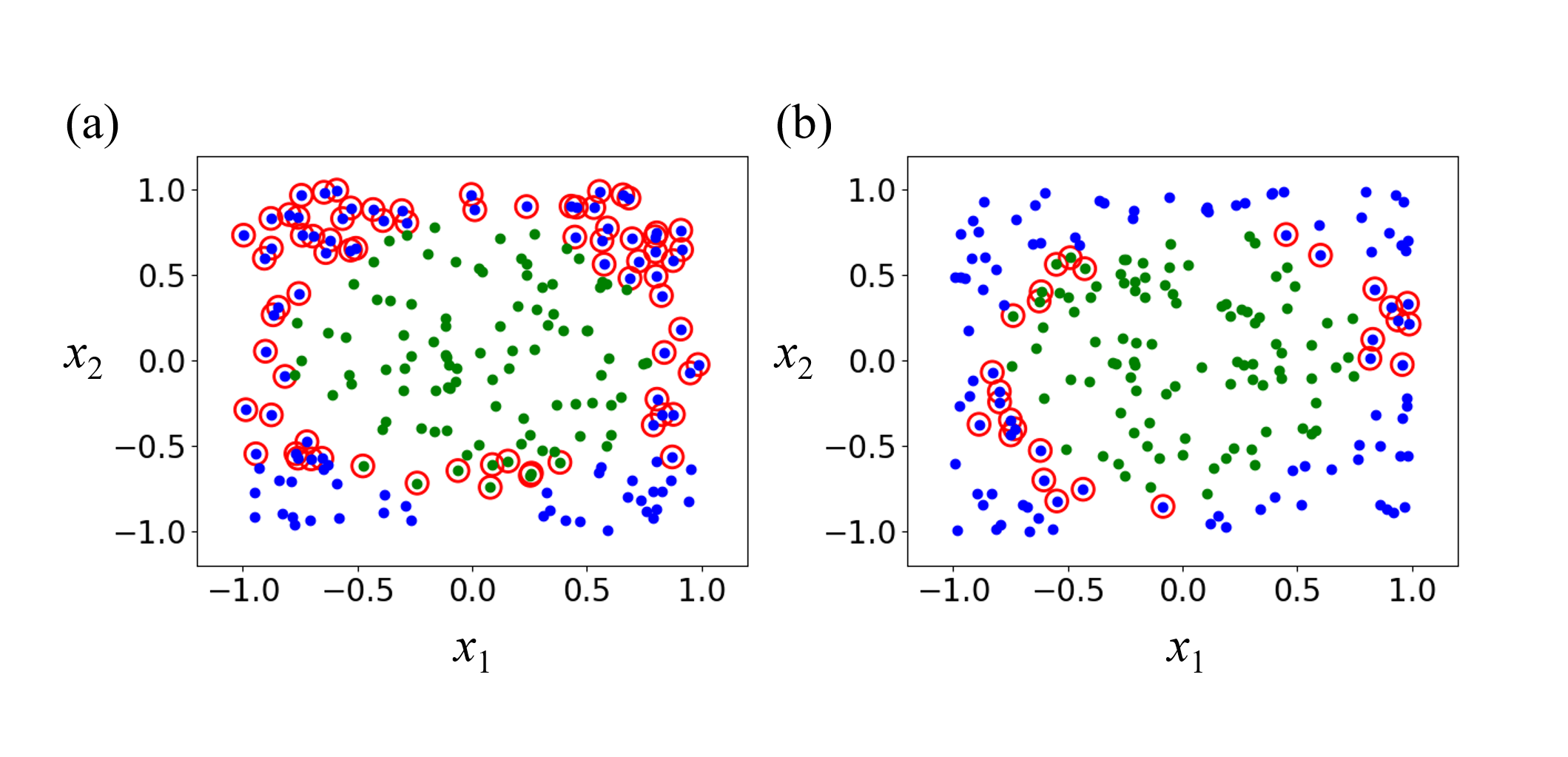}
    \caption{Classification result of 200 randomly selected samples for the circuits with classical and quantum control of causal order. The wrongly classified samples are marked with red circles. (a) With the classical control of causal order, the circuit is able to identify the lower edge part, the boundary is a line at about $x_2=-0.75$ and the accuracy increases to $62\%$. (b) With the quantum control of causal order, the accuracy increases to $86\%$. Only some sample points around the classification boundary are classified wrongly. This implies the learnt boundary becomes the correct circle boundary almost.}
    \label{claquan}
\end{figure}

The variational parameters makes up a 12-dimensional vector $[\phi,\theta,\omega,\{cr_{ij}\}_{i,j=1,2,3}]$. 
In classical $N$-switch, different orders exist simultaneously, but there is no coherence between them. Therefore, the observable is $ \sum_{\pi_i \in E}\ket{\pi_i}\bra{\pi_i}^{\alpha}  \otimes I^{c t_1 t_2 t_3} \otimes \sigma_z^t\in \mathcal{H}^{\alpha c t_1 t_2 t_3 t}$. The output is the linear combination of different orders.
\mnn{Here, we would like to point out that due to the redundancy of the ancillary system $\mathcal{H}^\alpha$, the final output differs slightly from the ideal one without the redundancy space. That is, with the final state $\ket{\psi_\pi}^\alpha=\sum_{i}c_{i}\ket{\pi_i}^\alpha$ of $\mathcal{H}^\alpha$ and the observable $ \sum_{\pi_i \in E}\ket{\pi_i}\bra{\pi_i}^{\alpha}$ measured on the effective space $E$, the ideal state of $\mathcal{H}^\alpha$ should be $\ket{\psi_\pi}^\alpha_{\text{ideal}}=\gamma
\sum_{\pi_i\in E}c_i\ket{\pi_i}^\alpha$ with a normalization factor $\gamma=\left(\sum_{\pi_i\in E}|c_i|^2\right)^{-1}$. Therefore, this difference only involves a positive scaling factor $\gamma$ and hence does not affect the predicted label or the accuracy of the ideal structure.}
\begin{figure}
    \centering
    \includegraphics[scale=0.4]{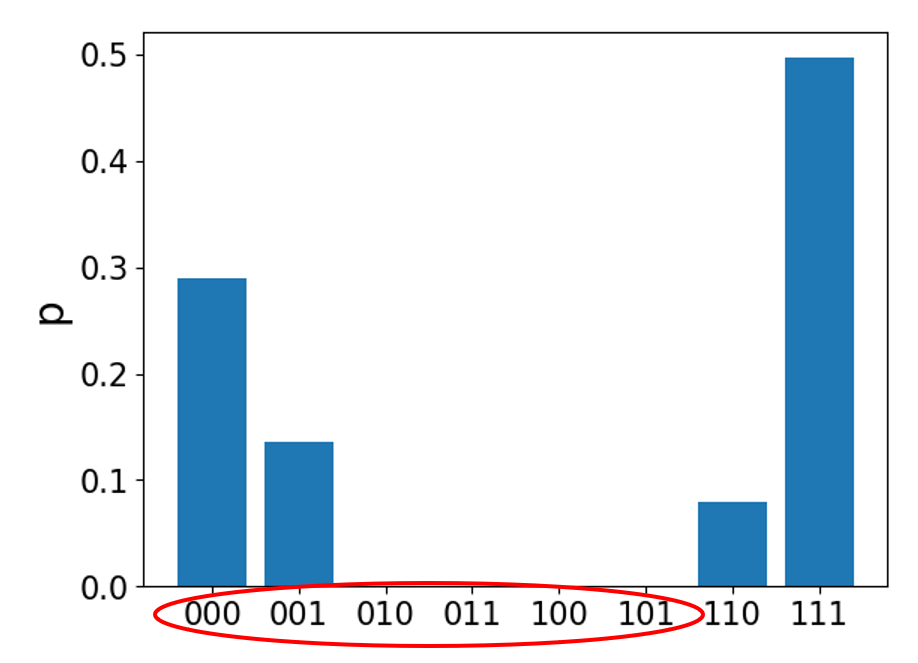}
    \caption{Possibilities of the final state based on classical $N$-switch. The first six bases represent different orders. Among these six bases, the dominant outputs are observed for $\ket{000}$ and $\ket{001}$, resulting in a final result that is a linear combination of these two orders.}
    \label{clas}
\end{figure}
With the classical control of causal order, the final accuracy significantly improves to $62\%$ with the optimal parameter vector $[1.7567, 2.2360, 0.7842, 0.9630, 0.2881, 0.7284, 1.8383,\\ 0.8536, -0.0387, 0.0011, 0.4796, 0.7081]$. The learnt classification boundary is about $x_2=-0.75$, i.e., the blue points in the lower edge part are classified correctly, as shown in Fig.~\ref{claquan}(a). This demonstrates that the new structure has successfully learned a significant portion of the target pattern. Thus, the enhanced expressivity of the circuit from classical control of causal order is confirmed. 
\begin{figure}
    \centering
    \includegraphics[scale=0.4]{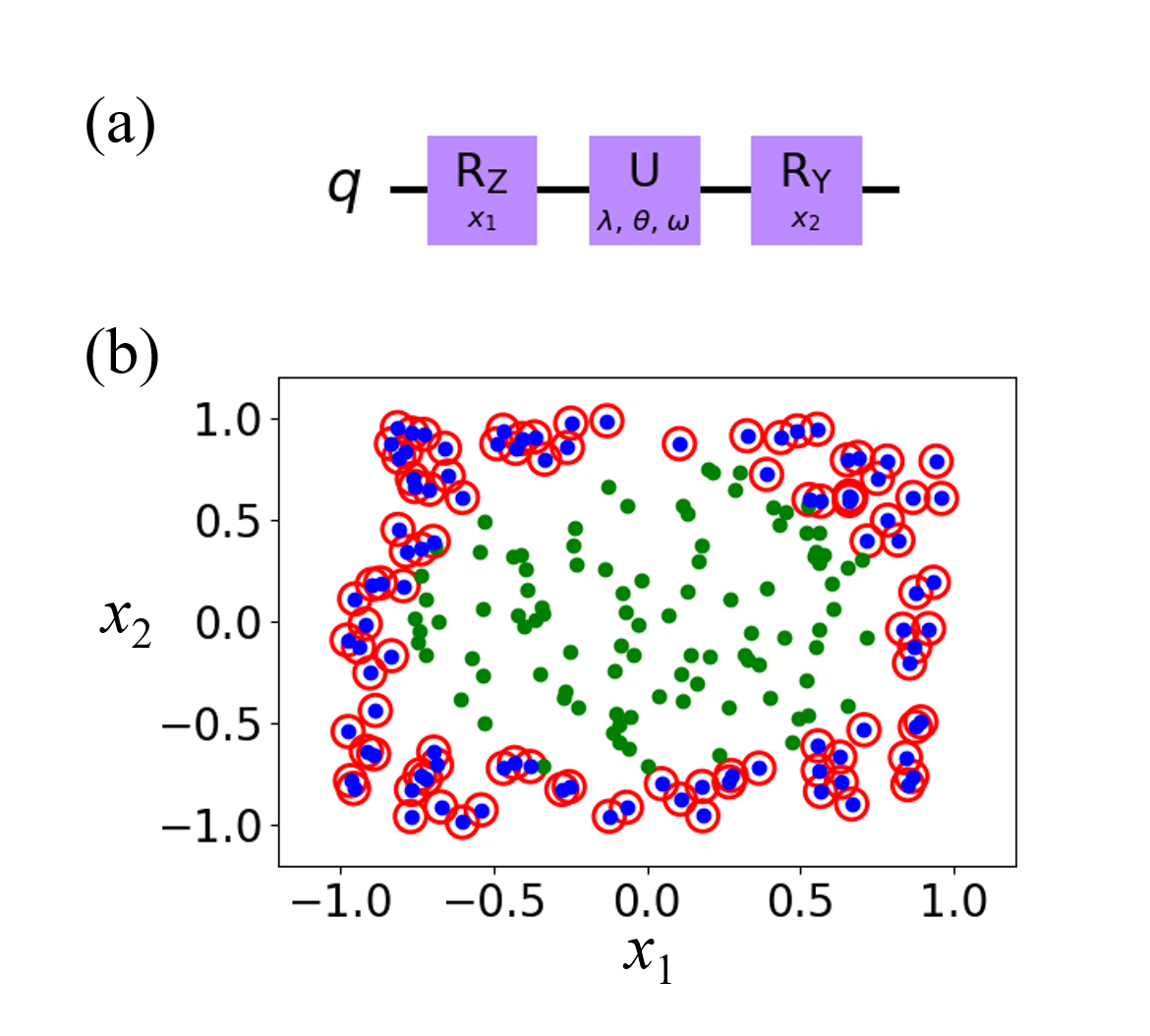}
    \caption{Circuit with order $[1,0,2]$ and the classification result for two dimensional binary classification task. (a) The circuit with order $[1,0,2]$. The overall unitary operator is $R_Y(x_2)U(\lambda, \theta, \omega)R_Z(x_1)$ (b) The
    classification result of 200 randomly selected samples. The wrongly classified samples are marked with red circles. The accuracy is $50\%$.}
    \label{fix1}
\end{figure}
As the output is the linear combination of different orders and the corresponding coefficients are proportional to the possibilities on the basic states, we project the final state $\ket{\psi}^\alpha$ to the bases
in the ancillary system. The possibility distribution is shown in Fig.~\ref{clas}. The result indicates that the optimal output using the classical $N$-switch occurs when combining the orders $[0,1,2]$ and $[1,0,2]$ mainly. In line with the explanation in the fixed order case, the circuit with the order $[0,1,2]$ fails to capture the underlying pattern. Similarly, the circuit with the order $[1,0,2]$ (shown in Fig.~\ref{fix1}) also struggles to recognize the distribution pattern with the optimal parameters $[1.6623, 1.5971, 0.7883]$. However, a linear combination of the two orders demonstrates an improved learning capability, capturing a significant portion of the target pattern. This improvement validates the efficacy of the classical control of causal order.

If we integrate the quantum control of causal order into the learning model, the accuracy can be further improved to a higher level.
To that end, we define our observable as $\sum_{\pi_i,\pi_j \in \alpha_E}\ket{\pi_i}\bra{\pi_j}^\alpha\otimes I^{c t_1 t_2 t_3}\otimes \sigma_z \in \mathcal{H}^{\alpha c t_1 t_2 t_3 t}$ which, by construction, can exploit the interference between different casual orders.  \mnn{In this setup, the term $\ket{\pi_i}\bra{\pi_j}^\alpha$ with $i\neq j$ represents the coherence of different order $\pi_i$ and $\pi_j$, a term absent from classical control. The diagonal part of $\ket{\pi_i}\bra{\pi_i}^\alpha$, akin to the classical $N$-switch scenario, is not related to the coherence between different orders and thus only representing a mixture of different orders.}
With the optimal parameters $[1.6628, 1.6813, 1.3400, 1.4326, 1.5209, 0.0921, 0.5260,\\ 0.9382, 1.2003, 1.5877, -0.3781, -0.5879] $ from the training, the accuracy improves further to $86\%$. From the distribution of the classification result in Fig.~\ref{claquan}(b), it can be observed that a few points near the boundary are misclassified. However, the learned boundary approaches the actual circular boundary quite closely. As a result, this model has successfully learned the underlying pattern to a significant extent.
\begin{figure}
    \centering
    \includegraphics[scale=0.4]{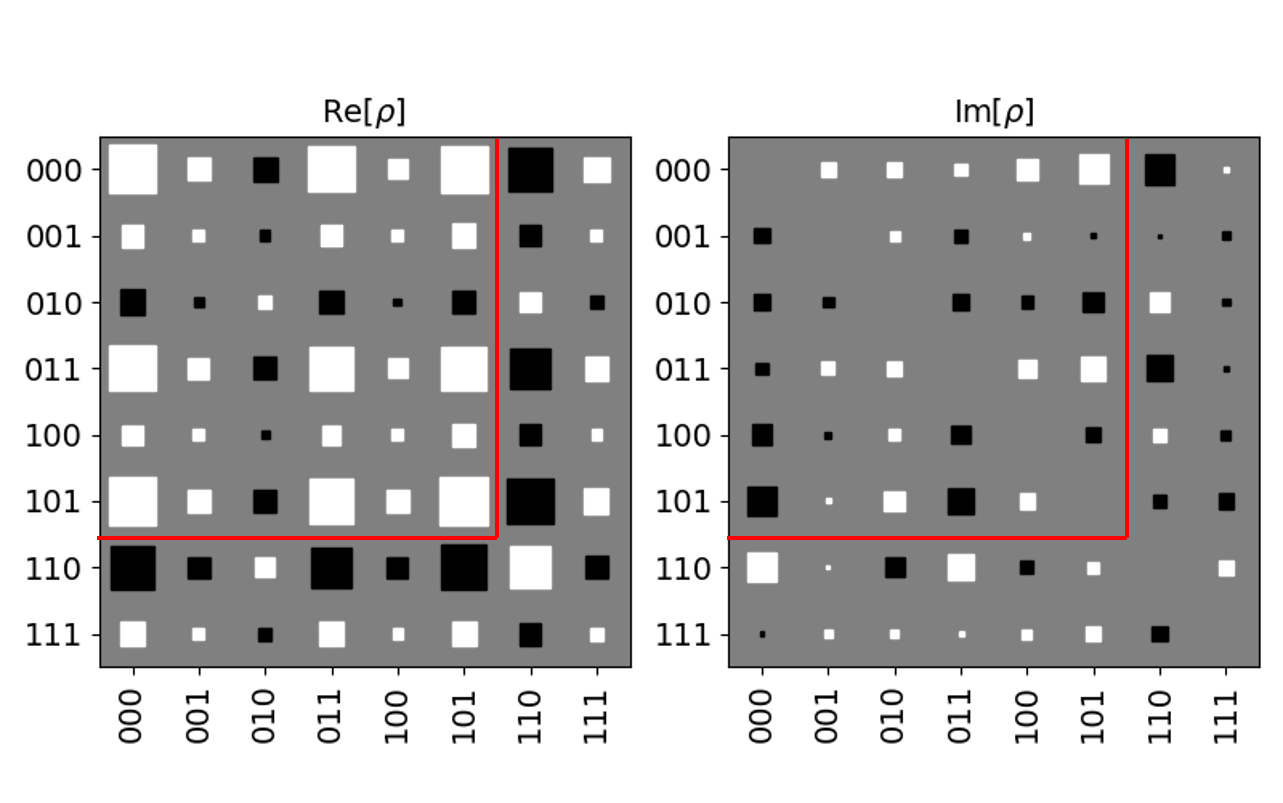}
    \caption{Real and imaginary parts of the ancillary system state matrix using quantum 3-switch. The color of each square indicates the sign of the corresponding element, with white representing positive values and black indicating negative values. The size of each square reflects the absolute value. The first six bases (highlighted by red lines) are of particular interest. The analysis reveals a strong coherence among the orders $[0,1,2]$, $[1,2,0]$, and $[2,1,0]$.}
    \label{quanod}
\end{figure}

To investigate the coherence among different orders, the state matrix $\rho$ of the ancillary system is presented in Fig~\ref{quanod}. The left-hand side displays the real part of the state matrix, while the right-hand side shows the imaginary part. The color of each square indicates the sign of the corresponding matrix element, with white representing positive values and black indicating negative values. The size of each square reflects the absolute value of the element. Since the last two bases, $\ket{110}$ and $\ket{111}$, serve as redundant parts with no specific order, our focus is solely on the first six bases. Notably, the bases $\ket{000}$, $\ket{011}$, and $\ket{101}$ dominate the matrix, suggesting that the orders $[0,1,2]$, $[1,2,0]$, and $[2,1,0]$, along with their coherence, play a decisive role in determining the output.
Upon revisiting the performance with 2 re-uploading layers with fixed order, we observe an improvement to an accuracy  $78\%$, albeit not surpassing the results obtained with quantum control of causal order. Although this setup incorporates more gates (six gates in this case), the comparison with the quantum control case suggests that while expanding the spectrum range with more layers, the flexibility of the respective coefficients may still constrain our model from achieving superior learning capabilities. This observation implies that, in many cases, the frequency spectrum alone may be adequate, but the restricted flexibility of the coefficients hinders the improvements in model performance.

\begin{figure}
    \centering
    \includegraphics[scale=0.35]{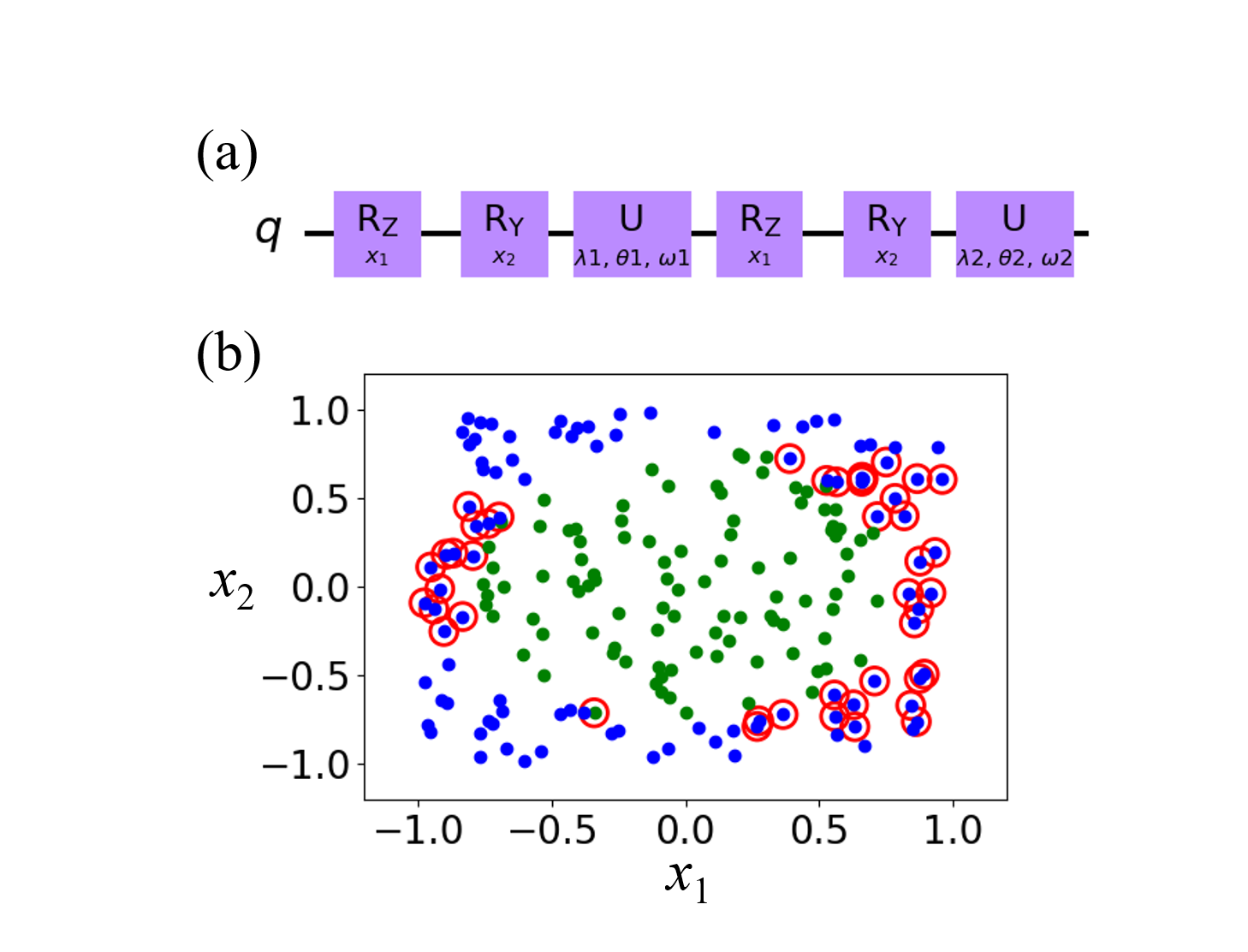}
    \caption{ Circuit with a fixed order and 2 re-uploading layers, along with its classification result. The performance demonstrates improvement compared to the single layer case. Correct classification is observed for the upper and lower parts of the blue points, resulting in an accuracy of $78\%$.}
    \label{fix2}
\end{figure}

Based on the above results, we can observe that as the control of causal order becomes more complex, the expressivity of a quantum circuit improves significantly.
By controlling the causal order, we can manipulate the flow of information and the quantum coherence within the circuit, allowing for more intricate and sophisticated computations.
Therefore, the control of causal order plays a crucial role in harnessing the full potential of a quantum circuit, enabling it to tackle more intricate tasks and enhance its learning ability.

\section{conclusion}

In this paper, we have investigated enhanced learning ability achieved with classical order and indefinite causal order, in contrast to the prevailing fixed-order structure in current quantum machine learning models. Through an analysis of the output function, we have elucidated that classical control of causal order only uses a linear combination of different orders. By contrast, indefinite causal order introduces coherence between different orders, resulting in more complex and coherent coefficients. These improvements are further confirmed by two examples, the output function of a 2-switch structure and the performance of a 3-switch structure in a binary classification task. 
With the improvement from quantum causal order having been demonstrated through simulations on a quantum circuit in this paper, its actual performance on a real quantum device with different causal orders needs to be confirmed. Additionally, beyond the $N$-switch structure, the dynamic quantum causal order structure \cite{wechs2021quantum} is more intricate, and hence exploring the effects of this structure should be a promising avenue for future studies.

%

\end{document}